Title: Giant adiabatic spin torque in magnetic tunnel junctions with a hybrid free layer structure

Authors: Xiaofeng Yao[1], Yisong Zhang[1], Andrew Lyle[1], Roger Malmhall[2], Rajiv Ranjan[2], Lei Lu[3], Mingzhong Wu[3], Hao Wang[1], Ying Jing[1], and Jian-Ping Wang[*1]

[1]: Department of Electrical and Computer Engineering,

the Center for Micromagnetics and Information Technologies (MINT),

University of Minnesota, MN 55455, USA

[2]: Avalanche Technology, Fremont, CA 94538, USA

[3]: Department of Physics, Colorado State University, CO 80523, USA

*: Corresponding author: jpwang@umn.edu; Tel: 612-625-9509

Author contributions:

X.F.Y. designed the experiments, patterned and fabricated the devices, carried out the transport measurements, analyzed the data, and drafted the paper. Y.S.Z. and X.F.Y. developed the macrospin model and performed the numerical simulation. R.M. and R.R. provided some of the magnetic tunnel junction wafers in-kind according to the design from J.P.W. and X.F.Y. Y.S.Z. and A.L. helped analyze the data. L.L. and M.Z.W. carried out the damping constant measurement. H.W. and Y.J. performed the transmission electron microscopy measurement and analysis. J.P.W. planned the experiments, designed the hybrid magnetic tunnel junctions by inserting a high spin reflection layer adjacent to the free layer, led the project, and finalized the paper. All authors were involved in the interpretation of the results and the preparation of the paper.

Abstract

The discovery of spin torque transfer (STT)[1,2,3] has lead to a significant advance in the development of spintronic devices. Novel structures and materials have been studied in order to improve the performance of the magnetic tunnel junctions (MTJs) performances and understand the fundamental physics in spin torque transfer[4,5,6,7,8,9,10]. The adiabatic spin torque effect, which is due to the spatial non-uniformity of magnetic properties, has been predicted in theory[11,12,13,14,15] and demonstrated experimentally in magnetic nanowires[16,17,18,19]. However, this important spin torque has been rarely concerned in the magnetic tunnel junctions (MTJ) because of its extremely weak effect in conventional MTJs. This paper reports for the first time a giant adiabatic spin torque in MTJ devices with a hybrid free layer structure. The generation of the giant adiabatic spin torque was realized through the introduction of a spatial magnetic non-uniformity in a hybrid free layer along the current direction. It is observed that the giant adiabatic spin torque can substantially promote the current-induced switching process in the MTJ devices: the adiabatic spin torque can be larger than the in-plane spin torque, which allows for the switching with a single-polar current under different bias fields. Moreover, the adiabatic spin torque, which is proportional to the level of spatial non-uniformity, increases during the switching process. The observed effects are confirmed by numerical simulations. These results have far-reaching implications for the future of high-density STT-MRAM devices.

Magnetic tunnel junctions (MTJ), which can have extremely large magnetoresistance, are one of the most important building blocks for spintronics devices, in particular, spin torque transfer magnetic random access memory (STT-MRAM) devices. The operation of the spin torque transfer is realized through the interactions between spin-polarized conduction electrons and local magnetic moments. The concept of the spin torque transfer was first proposed by Slonczewski[1] and Berger[2] in 1996. In conventional MTJs, which have one free layer and one pinned layer, there are two different types of spin torques that can be transferred:

$$\tau = -\beta_{IT}\vec{M}_f \times (\vec{M}_f \times \vec{M}_p) - \beta_{PT}\vec{M}_f \times \vec{M}_p \tag{1}$$

where $\vec{M}_f$ and $\vec{M}_p$ are the magnetizations for the free and pinned layers, respectively, and $\beta_{IT}$ and $\beta_{PT}$ are two coefficients that depends mainly on spin polarization ratio and saturation magnetization of the magnetic layers. The first term is often called the in-plane spin torque. The second term is called the perpendicular spin torque, which was not explored until very recently[20,21,22,23].

The concepts described above are based on the assumption that the free layer has a spatially uniform magnetization along the direction of the current. In novel MTJ structures, however, very often two or more magnetic thin films are used as a composite free layer. This composite configuration results in a spatial non-uniformity of the magnetic properties along the current direction, which leads to additional spin torques. In conventional MTJ devices, those non-uniformity induced spin torques are very weak and, therefore, have been rarely concerned. In contrast, there have been a number of previous works that studied the effect of such spin torques on the dynamics of the current-induced domain wall motion in magnetic nanowires[11,12,13,14,15,16,17,18,19]. The non-uniformity associated spin torques fall into two different categories: adiabatic and non-adiabatic. The adiabatic spin torque corresponds to the situation where the spatial variation of the magnetization is gradual and the spins of the conduction electrons can follow the direction of the local moments. The non-adiabatic spin torque is the case when there is a sharp change in the magnetization and the spins of the conduction electrons cannot follow the direction of the local moments. One can describe these torques as[15]

$$\vec{\tau} = -b_{AT}\vec{M}_f \times (\vec{M}_f \times \frac{\partial \vec{M}}{\partial z}) - b_{NT}(\vec{M}_f \times \frac{\partial \vec{M}}{\partial z}), \qquad (2)$$

where the first and second term represent the adiabatic and non-adiabatic spin torque, respectively, $\vec{M}$ is the magnetization within the non-uniform region, $z$ is a space variable in the direction of the current, and $b_{AT}$ and $b_{NT}$ are the coefficients that depend on the spin polarization ratio and the saturation magnetization of magnetic layers of the testing devices. It is worth to note that in this paper we follow the use of the names of adiabatic and non-adiabatic spin torques due to a spatial magnetic non-uniformity of the magnetic properties in magnetic layers, which are typically used in the studies of current-induced domain wall motion.

Based on the above discussion, there are four types of spin torques that can be involved in the current-induced magnetization switching process in MTJ devices. The in-plane and adiabatic spin torques act along the same axis within the layer plane, while perpendicular and non-adiabatic spin torques have directions perpendicular to the layer plane. When there are two or more spin torques (with the same type, or different types) along the same axis, the net spin torque exerted on the free layer will be augmented or diminished, depending on the relative orientations of the magnetizations in these magnetic layers, and the relative position of these magnetic layers to the free layer. Many works have been done to study on the enhancement or reduction of two in-plane spin torques in giant magnetoresistance (GMR) or MTJ devices with dual pinned structures[4,24,25,26] or synthetic antiferromagnetic structures[27,28]. However, there is little study on the effect of the adiabatic spin torque in MTJ devices. Recently, Kim's paper[29] introduces the adiabatic spin torque term in their micromagnetic simulation of MTJ devices with polarization enhancement layers. However, there is no experimental demonstration of the adiabatic spin torque effect in MTJ devices. More fundamental studies in both theory and experiment on adiabatic spin torques in MTJ devices are required to improve the device performances.

This paper reports for the first time a giant adiabatic spin torque in MTJ devices with a hybrid free layer structure. The generation of the giant adiabatic spin torque was realized through the introduction of the spatial magnetic non-uniformity in a hybrid free layer structure. Such a giant adiabatic spin torque can substantially promote the current-induced switching process in the MTJ device: the adiabatic spin torque can be larger than the in-plane spin torque, which allows for the switching with a single-polar current under different bias fields. These results have far-reaching implications for the future of high-density STT-MRAM devices.

The key to realize the giant adiabatic spin torque is the introduction of the spatial magnetic non-uniformity in the hybrid free layer structure by inserting a high spin reflection layer to the free layer side of a conventional MTJ structure. The high spin reflection layer needs to satisfy three criteria to generate the giant adiabatic spin torque. First, the spin reflection layer should have a relatively large saturation magnetization compared to the free layer. In general, the larger the magnetization is, the higher the spin polarization ratio and the amplitude of the adiabatic spin torque are. Second, the exchange coupling between the spin reflection layer and the free layer should be ferromagnetic and relatively weak. Third, the spin reflection layer should have stronger magnetic anisotropy than the free layer. The latter two criteria ensure that the free layer can be switched independently, while the spin reflection layer will not be switched.

It is worth mentioning that, for the device structure discussed below, the perpendicular and non-adiabatic spin torques are much smaller than the in-plane torque and the adiabatic torque, respectively, and, therefore, can be neglected. As a result, only the in-plane spin torque and the adiabatic spin torque are responsible for the magnetization switching in the free layer. Thus the total torque can be written as

$$\tau = -\beta_{IT} \vec{M}_f \times (\vec{M}_f \times \vec{M}_p) - b_{AT} \vec{M}_f \times (\vec{M}_f \times \frac{\partial \vec{M}}{\partial z}), \qquad (3)$$

For the data presented below, the hybrid free layer structure is composed of a $Co_{60}Fe_{20}B_{20}$ layer and a thin FeSiO layer. The $Co_{60}Fe_{20}B_{20}$ layer is the free layer, since its magnetization direction controls the MTJ device resistance. The thin FeSiO layer acts as the high spin reflection (HSR) layer. The Fe in the FeSiO layer provides large saturation magnetization and high spin polarization ratio, while the $SiO_2$ dopant ensures the high anisotropy of the film and the weak exchange coupling between the FeSiO film and the CoFeB free layer. The fixed layer is composed by a synthetic antiferromagnetic structure. The MTJ devices have a multilayered structure as follows: Si substrate / $SiO_2$ (100 nm) / bottom lead / PtMn (20 nm) / $Co_{70}Fe_{30}$ (2.5 nm) / Ru (0.85 nm) / $Co_{60}Fe_{20}B_{20}$ (2.4 nm) / MgO (1.0 nm) / $Co_{60}Fe_{20}B_{20}$ (2.0 nm) / FeSiO (1.5 nm) / top lead. The devices were patterned into an ellipse shape by the use of e-beam lithography followed by an ion milling process. The lateral dimensions are either 130 nm by 162 nm or 149nm by 185nm.

The transport measurements were carried out under different currents and bias magnetic fields. The positive current is defined as the electrons flowing from the fixed layer to the free layer. The direction of positive magnetic fields is defined as the direction opposite to the magnetization in the fixed layer. The tunnel magnetoresistance ratio (TMR) ranges from approximately 100% to 130%. The production of the resistance and area (RA) values for the parallel state is 10~12 $\Omega \cdot \mu m^2$. The bias field discussed below is defined as: $H_{bias} = (H_{applied} - H_{offset})/H_c$, where $H_{applied}$, $H_{offset}$, $H_c$ are the applied magnetic field, the offset field, and the coercive field, respectively. Note that the offset field originates from the dipole field from the fixed layer. The orange peel coupling from the interface roughness may also contribute the field offset.

Figure 1 (a) and (b) shows the transport properties of the MTJ devices with hybrid free layer structures. The current induced switching from antiparallel (AP) state to parallel (P) state under different bias fields is shown in Fig.1(a). The MTJ device was first set to AP state as shown in the schematic drawing of Fig.1(c). A negative bias field is applied, which lowers the energy barrier of the CoFeB free layer between P and AP states. The positive current is applied with the electrons first polarized by the fixed layer. The majority spins have opposite direction to the CoFeB free layer. Therefore, an in-plane spin torque is exerted on the CoFeB free layer to switch it to the parallel state, as shown in the drawing of Fig.1(e). As discussed previously, the CoFeB free layer has smaller anisotropy than FeSiO HSR layer and weak exchange coupling with FeSiO HSR layer. Therefore, CoFeB free layer switches first, while FeSiO remains in its initial state. As the CoFeB free layer switches, the spatial magnetic non-uniformity along the current direction in the hybrid free layer structure increases consequently, which enhances the adiabatic spin torque, through the reflected majority spins. The adiabatic spin torque has the same direction with the in-plane spin torque and helps switch CoFeB free layer to the P state as shown in the drawing of Fig.1(e). When the bias field amplitude decreases, the critical current increases as shown in Fig.1(a). The switching current under the different bias field in Fig.1(a) and (b) is the average value based on multiple measurements. The AP to P switching in MTJs with the hybrid free layer structure is similar to the conventional MTJ devices, which use positive current to switch the free layer from AP to P.

Interestingly, the P to AP switching in the MTJs with hybrid free layer structures is much different from the conventional MTJs. Fig. 1(b) shows that the same positive current can also switch the MTJ devices from P to AP state under different bias fields, which cannot be realized with conventional MTJ devices. The MTJ device is pre-set to parallel state as shown in the schematic drawing in Fig.1(d). The positive bias field is applied, which lowers the energy barrier of CoFeB free layer between P and AP states. The same positive current is applied through MTJ devices with electrons flowing from the fixed layer to the CoFeB free layer. The polarized electrons have majority spins with the same direction to the CoFeB free layer and the FeSiO HSR layer. Thus the in-plane spin torque, which is exerted by the majority spins coming from the fixed layer, helps keep the CoFeB free layer parallel to the fixed layer as shown in Fig.1(f). However, due to the spatial magnetic non-uniformity in the hybrid free layer structure along the current direction, the minority spins that flow through the interface region and the FeSiO HSR layer, are scattered back to the CoFeB free layer. These reflected minority spins exert the adiabatic spin torque, which has opposite direction to the in-plane spin torque as shown in the drawing of Fig.1(f). Moreover, during the switching of the CoFeB free layer, the FeSiO HSR layer remains in its initial state as discussed early in this paper. Therefore, the amplitude of adiabatic spin torque increases with the switching of the CoFeB free layer since it is proportional to the level of the spatial non-uniformity in the hybrid free layer. This kind of dynamic spin torque during the switching process due to the spatial non-uniformity of the magnetization is a unique property in the MTJ devices with our proposed hybrid free layer structure, which is different from the MTJ devices with the dual pinned structures in previous reported works[4,24,25,26]. As shown in Fig.1(f), the adiabatic spin torque competes with the sum of the in-plane spin torque and the damping coming from the effective field. When the applied current is larger than a critical value, the adiabatic spin torque is larger than the sum of the in-plane spin torque and the damping, which induces the unique P to AP switching as shown in Fig.1(b). This implies that the adiabatic spin torque increases with the current faster than the in-plane spin torque. This unique P to AP switching with positive current has been observed in a wide range of the bias field from 0.710 to 0.903. As the decrease of the bias field, the effective field, which is dominated by the anisotropy field, becomes larger which results in the increase of the contribution from the damping. Since the adiabatic spin torque competes with the sum of the in-plane spin torque and the damping, the excess of the adiabatic spin torque compared to the in-plane spin torque should become larger with the decrease of the bias field. As shown in the experimental results, when the bias field is reduced to 70% of the coercive field (switching field), the adiabatic spin torque can still be large enough to realize the unique P to AP switching. The final AP state of the MTJ devices is a stable state within the applied current range, and is also stable when the applied current is removed.

The giant adiabatic spin torque effect leads to the single-polar current switching in MTJ devices with hybrid free layer structures under different bias fields. As shown in Fig.2, by tuning the bias field, the same amplitude positive current can switch the MTJ device with a hybrid free layer structure in both directions (from P to AP and from AP to P). The switching in both directions is a sharp transition without any intermediate state. The same transport behaviour can be well repeated in different MTJ devices with two different sizes. The above unique P to AP switching is believed to result from the giant adiabatic spin torque generated by the spatial magnetic non-uniformity in the hybrid free layer structure. At the same time, it is important to exclude any thermal heating effect for such unique P to AP switching. If the thermal heating due to the large current passing through the MTJ device is the cause, a negative current with the same or even smaller amplitude should also induce such P to AP switching, as reported in ref. 24. However, the experimental result shows that even when a negative current three times larger than the positive switching current, the MTJ device remains in P state, which is shown in the Fig.3s(b) of the supplementary information. Therefore the unique P to AP switching with a positive current is not due to the thermal heating effect.

To gain more insight, the FeSiO HSR layer was analyzed to confirm that it indeed satisfied the three criteria to realize giant adiabatic spin torque as mentioned earlier in this paper. First, the HSR layer should have high saturation magnetization ($M_s$) and/or high spin polarization. The X-ray photoelectron spectroscopy (XPS) shows that Fe is the dominant phase in the FeSiO thin film, and there is no $Fe_2O_3$ peak in the Fe 2p3 XPS spectrum. Si has two phases in FeSiO thin film, silicon oxide and iron Silicide or silicon, as shown in Fig.1s of the supplementary information. The volume fraction of Fe, $SiO_2$ and (Fe)Si is approximate 70%, 15% and 15%, respectively. Based on the TEM images shown in Fig.4s of the supplementary information, the Fe and/or Fe(Si) grains are surrounded by $SiO_2$ grain boundaries. Besides the dominant phase of Fe, the $Fe_3Si$ phase is also believed to exist in FeSiO film. Both Fe and $Fe_3Si$[30] have high spin polarization ratio, which enhances the adiabatic spin torque during the current-induced switching process. The second criterion for a proper HSR layer is that there is relatively weak coupling between FeSiO and CoFeB layers. This criterion is satisfied by using $SiO_x$ in FeSiO layer, which diffuses to the grain boundaries and also the interface between FeSiO and CoFeB layers during the post-annealing process. The diffusion of $SiO_x$ reduces the interlayer exchange coupling between FeSiO and CoFeB layers. The third criterion is that FeSiO layer should have higher anisotropy than CoFeB layer so that the CoFeB switches independently with a lower switching current. The magnetic anisotropy constants of the annealed FeSiO and CoFeB thin films were estimated, with the Law of Approach to Saturation (LATS) based on the initial magnetization curves, as 4.0 and $2.0 \times 10^4$ erg/cm$^3$, respectively. The increase of the magnetic anisotropy constant in FeSiO thin film has been studied by other groups[31] and may be due to changes of particle size and percolation effect. Moreover, the damping constant for FeSiO and CoFeB thin films was measured using the ferromagnetic resonance (FMR) method. The effective damping constant of FeSiO and CoFeB films are 0.015 and 0.008, respectively, as shown in Fig.2s of the supplementary information. Therefore, due to the lower anisotropy constant and damping constant, the CoFeB free layer has a lower critical switching current than the FeSiO HSR layer[32]. Thus based on the above analysis, the FeSiO does satisfy all three of the required criteria to realize the giant adiabatic spin torque.

In order to confirm the experimental observations, numerical analysis based on a macrospin model has been done. The motion of the CoFeB free layer moment is described by modified Landau-Lifshitz-Gilbert (LLG) equation, which includes in-plane and adiabatic spin torque terms as shown in the following:

$$\frac{\partial \vec{M}_2}{\partial t} = -\gamma \vec{M}_2 \times \vec{H}_{eff2} - \alpha^* \vec{M}_2 \times (\vec{M}_2 \times \vec{H}_{eff2}) - \beta_{IT2} \vec{M}_2 \times (\vec{M}_2 \times \vec{M}_3) + b_{AT2} \vec{M}_2 \times (\vec{M}_2 \times \frac{\partial \vec{M}_{1,2}}{\partial z})$$

For the FeSiO HSR layer, there is only an adiabatic spin torque term in LLG equation, which is: $\frac{\partial \vec{M}_1}{\partial t} = -\gamma \vec{M}_1 \times \vec{H}_{eff1} - \alpha^* \vec{M}_1 \times (\vec{M}_1 \times \vec{H}_{eff1}) + b_{AT1} \vec{M}_1 \times (\vec{M}_1 \times \frac{\partial \vec{M}_{1,2}}{\partial z})$. The subscript number 1, 2 and 3 represent the FeSiO HSR layer, CoFeB free layer and CoFeB fixed layer, respectively. $\alpha_{1,2}^* = \frac{\gamma \alpha_{1,2}}{(1+\alpha_{1,2}^2) M_{s1,2}}$, $\beta_{IT2} = \frac{\mu_B J P_3}{e d_2 M^2_{s2} M_{s3}}$,

$b_{AT1,2} = \frac{\mu_B J P_{2,1}}{e M^2_{s1,2} M_{s2,1}}$ are coefficients of the damping, in-plane spin torque and adiabatic spin torque, respectively. γ is the gyromagnetic ratio, α is the Gilbert damping constant, $M_s$ is the saturation magnetization, P is spin polarization ratio, $\mu_B$ is the Bohr magneton, e is electron charge, and J is the current density, d is layer thickness. The X,Y and Z axes are the in-plane easy-axis, hard-axis and out-of-plane direction (current direction), respectively. Based on the macrospin approximation, $\frac{\partial \vec{M}_{1,2}}{\partial z} \approx \frac{\vec{M}_1 - \vec{M}_2}{d^*}$, where

$d^* = \frac{(d_1 + d_2)}{2}$ represents the distance of the magnetization change in the interface region of the CoFeB free layer and the FeSiO HSR layer. The effective field includes the external field, magnetic anisotropy field, demagnetic field and interlayer coupling field. The saturation magnetization, magnetic anisotropy constant and damping constant used in the simulation are based on the experimental results. The spin polarization ratio of FeSiO and CoFeB is set to 0.55 and 0.40, respectively. Fig.3(a) and (b) show the simulation result for AP to P switching with a positive current and a negative bias field (-0.903), which is the same as the experimental setting. The critical switching current density ($J_c$) for CoFeB and FeSiO layers are $0.8 \times 10^7$ A/cm$^2$ and $6.1 \times 10^7$ A/cm$^2$, respectively. The simulation value of $J_c$ is about four times larger than the experimental $J_c$ value, which is attributed to the fact that the simulation is performed at zero Kelvin and no temperature effect is considered.

The unique P to AP switching with a positive current and a positive bias field (0.903) is also obtained in the simulation as shown in Fig.3(c). The switching current for CoFeB free layer is $4.4 \times 10^7$ A/cm$^2$. This simulation result agrees with the experimental observation of the unique P to AP switch. When the adiabatic spin torque term is removed from the LLG equations, such unique P to AP switch cannot be obtained with any current value, which indicates that the giant adiabatic spin torque plays an indispensable role for the unique P to AP switching. At the CoFeB critical switching current, the FeSiO HSR layer does not switch as shown in Fig.3(d). Since the device resistance is dominated by the relative alignment of the two magnetic layers adjacent to the tunneling barrier, the contribution from the FeSiO HSR layer orientation to the device resistance can be neglected. When a larger current is applied, the FeSiO HSR layer switches to AP state, while the CoFeB free layer switches back to P state as shown in Fig.3(e) and (f). However, when the applied current is larger than $10 \times 10^7$ A/cm$^2$, both CoFeB free layer and FeSiO HSR layer prefer P state as shown in Fig.3(g) and (h). This behavior was not observed in the experiment since the current is higher than the breakdown current of the MTJ devices. Fig.3(c)~(h) shows that the unique P to AP switching with a positive current only happens within a window of applied current. Such an operation window for the unique P to AP switching depends on the difference of the spin polarization ratio and the interlayer exchange coupling between CoFeB free layer and FeSiO HSR layer, which was discussed in the first and second criteria of the HSR layer earlier. The simulations show that such an operation window is only obtained with a weak interlayer coupling between the FeSiO HSR layer and the CoFeB free layer. When the interlayer coupling is strong, CoFeB free layer switches together with FeSiO layer, and P to AP switch cannot happen with any positive current.

Fig.4 summaries the possible current-field (I-H) phase diagram for the switching of MTJ devices. The blue dashed lines indicate the magnetic reversal phase boundaries in conventional MTJ devices[5,10,33], where P is the parallel state, AP is the antiparallel state, and OS shows the oscillation region. The black triangle and purple square dots are experimental data of the MTJ devices with hybrid free layer structures. The AP to P switch of the hybrid MTJ devices follows the phase boundaries of the convention MTJ devices. However, the unique P to AP switch in the MTJ devices with hybrid free layer structures forms a special region in the phase diagram, which is named as zone Z shown in the purple area in Fig.4.

In summary, we proposed and fabricated a new type of MTJ devices that has a hybrid free layer structure. The hybrid structure introduces a spatial magnetic non-uniformity along the current direction, which exerts an adiabatic spin torque on the free layer. Through the optimization of the properties of the hybrid structure, we demonstrated an adiabatic spin torque that is surprisingly larger than the conventional in-plane spin torque. Such a giant adiabatic torque can plays a critical role in the switching of the magnetization in the free layer. In particular, this torque allows for both the AP to P and P to AP switching operations with a *single-polar current* under different bias fields. In addition, the experimental results were confirmed by numerical simulations that were based on a macrospin model. Our finding may open a door for the development of new MTJ-based STT devices.

Methods:

The MTJ multilayers were deposited by a sputtering method under Ar environment and ambient temperature. The full structure of MTJ is: Si substrate / $SiO_2$ (100 nm) / bottom lead / PtMn (20 nm) / $CoFe_{30}$ (2.5 nm) / Ru (0.85 nm) / $Co_{60}Fe_{20}B_{20}$ (2.4 nm) / MgO (1.0 nm) / $Co_{60}Fe_{20}B_{20}$ (2.0 nm) / FeSiO (1.5 nm) / top lead. A Synthetic antiferromagnetic structure is used as the pinned layer. The 2.0 nm CoFeB layer is the free layer and the 1.5 nm FeSiO layer is the HSR layer. After the deposition, the MTJ sample was annealed at 310ºC for 2 hrs in a 1 Tesla magnetic field.

For the device fabrication, E-beam lithography was used to pattern MTJ pillars and photo-lithography was used for bottom lead and top lead patterning. The MTJ devices have two different sizes with elliptical shape. The average lateral dimensions are 130 nm by 162 nm or 149nm by 185nm. The TMR ratio is 100%~130% and RA value at parallel state is 10~12 $\Omega um^2$.

For the device transport measurement, we use a four-point probe method. The positive current is defined as electrons flowing from the fixed layer to the free layer. Quasi-static current is applied with the pulse width of 0.5 second. The positive bias field is defined as the direction opposite to the fixed layer direction.

The analyzed CoFeB thin film and FeSiO thin film have structures of: Si sub/100 $SiO_2$ (100 nm)/ CoFeB (5 nm)/ Ta (2nm) and Si sub/ $SiO_2$ (100 nm)/ FeSiO (5 nm)/ Ta (2 nm), respectively. The magnetic property was measured using a vibration sample magnetometer (VSM); the composition and phase analysis was done by x-ray photoelectron spectroscopy (XPS); the damping constant was measured by a ferromagnetic resonance (FMR) method; the microstructure was measured by transmission electron microscopy (TEM).

Acknowledgement:

The authors would like to thank the partial support by National Science Foundation (NSF) ECCS (0702264), NSF Nano Fabrication Center at University of Minnesota, the Institute of Technology Characterization Facility at University of Minnesota, which receives partial support from NSF through the NNIN program, NSF Minnesota MRSEC Program.

[1] Slonczewski, J. C. Current-driven excitation of magnetic multilayers. *J. Magn. Magn. Mater.* 159, L1-L7 (1996).

[2] Berger, L. Emission of spin waves by a magnetic multilayer traversed by a current. *Phys. Rev. B* 54, 9353-9358 (1996).

[3] Katine, J. A., Albert, F. J., Buhrman, R. A., Myers, E. B. & Ralph, D. C. Current-driven magnetization reversal and spin-wave excitations in Co/Cu/Co pillars. *Phys. Rev. Lett.* 84, 3149-3152 (2000).

[4] Jiang, Y. et al. Substantial reduction of critical current for magnetization switching in an exchange-biased spin valve. *Nature Mater.* 3, 361-364 (2004).

[5] Mangin, S. et al. Current-induced magnetization reversal in nanopillars with perpendicular anisotropy. *Nature Mater.* 5, 210-215 (2006).

[6] Ravelosona, D. et al. Domain wall creation in nanostructures driven by a spin-polarized current. *Phys. Rev. Lett.* 96, 186604 (2006).

[7] Krivorotov, I. N. et al. Time-domain measurements of nanomagnet dynamics driven by spin-transfer torques. *Science* 307, 228-231 (2005).

[8] Houssameddine, D. et al. Spin-torque oscillator using a perpendicular polarizer and a planar free layer. *Nature Mater.* 6, 447-453 (2007).

[9] Meng, H., & Wang, J. -P. Composite free layer for high density magnetic random access memory with lower spin transfer current. *Appl. Phys. Lett.* 89, 152509 (2006).

[10] Ozatay, O. et al. Sidewall oxide effects on spin-torque- and magnetic-field-induced reversal characteristics of thin-film nanomagnets. *Nature Mater.* 7, 567-573 (2008).

[11] Berger, L. Low-field magnetoresistance and domain drag in ferromagnets. *J. Appl. Phys.* 49, 2156 (1978)

[12] Bazaliy, Ya. B., Jones, B. A., & Zhang, S. -C. Modification of the Landau-Lifshitz equation in the presence of a spin-polarized current in colossal- and giant-magnetoresistive materials. *Phys. Rev. B* 57, R3213-R3216 (1998).

[13] Tatara, G., & Kohno, H. Theory of current-driven domain wall motion: spin transfer versus momentum transfer. *Phys. Rev. Lett.* 92, 086601 (2004).

[14] Li, Z., & Zhang, S. Domain-wall dynamics driven by adiabatic spin-transfer torques. *Phys. Rev. B* 70, 024417 (2004).

[15] Zhang, S. & Li, Z. Roles of nonequilibrium conduction electrons on the magnetization dynamics of ferromagnets. *Phys. Rev. Lett.* 93, 127204 (2004).

[16] Thomas, L. et al. Oscillatory dependence of current-driven magnetic domain wall motion on current pulse length. *Nature* 443, 197-200 (2006).

[17] Hayashi, M., Thomas, L., Rettner, C., Moriya, R., & Parkin, S. S. P. Direct observation of the coherent precession of magnetic domain walls propagating along permalloy nanowires. *Nature Phys.* 3, 21-25 (2007).

[18] Thomas, L. et al. Resonant amplification of magnetic domain-wall motion by a train of current pulses. *Science* 315, 1553-1556 (2007).

[19] Kläui, M. et al. Direct observation of domain-wall configurations transformed by spin currents. *Phys. Rev. Lett.* 95, 026601 (2005).

[20] Li, Z. et al. Perpendicular spin torques in magnetic tunnel junctions. *Phys. Rev. Lett.* 100, 246602 (2008).

[21] Sankey, J. C. et al. Measurement of the spin-transfer-torque vector in magnetic tunnel junctions. *Nature Phys.* 4, 67-71 (2008).

[22] Deac, A. M. et al. Bias-driven high-power microwave emission from MgO-based tunnel magnetoresistance devices. *Nature Phys.* 4, 803-809 (2008).

[23] Kubota, H. et al. Quantitative measurement of voltage dependence of spin-transfer torque in MgO-based magnetic tunnel junctions. *Nature Phys.* 4, 37-41 (2008).

[24] Fuchs, G. D. et al. Adjustable spin torque in magnetic tunnel junctions with two fixed layers. *Appl. Phys. Lett.* 86, 152509 (2005).

[25] Huai, Y., Pakala, M., Diao, Z., & Ding, Y. Spin transfer switching current reduction in magnetic tunnel junction based dual spin filter structures. *Appl. Phys. Lett.* 87, 222510 (2005).

[26] Meng, H., Wang, J. G., & Wang, J.-P. Low critical current for spin transfer in magnetic tunnel junctions. *Appl. Phys. Lett.* 88, 082504 (2006).

[27] Hayakawa, J. et al. Current-induced magnetization switching in MgO barrier based magnetic tunnel junctions with CoFeB/Ru/CoFeB synthetic ferrimagnetic free layer. *Japan. J. Appl. Phys.* 45, L1057-L1060 (2006).

[28] Yen, C. -T. et al. Reduction in crtical current density for spin torque transfer switching with composite free layer. *Appl. Phys. Lett.* 93, 092504 (2008).

[29] Kim, W., Lee, T. D., & Lee, K. J. Current-induced flip-flop of magnetization in magnetic tunnel junction with perpendicular magnetic layers and polarization-enhancement layers. *Appl. Phys. Lett.* 93, 232506 (2008).

[30] Herfort, J., Schönherr, H.-P., & Ploog, K. H. Epitaxial growth of $Fe_3Si$/GaAs(001) hybrid structures. *Appl. Phys. Lett.* 83, 3912 (2003)

[31] Xiao, G. & Chien, C. L. Enhanced magnetic coercivity in magnetic granular solids. *J. Appl. Phys.* 63, 4252-4254 (1988).

[32] Sun, J. Z. Spin-current interaction with a monodomain magnetic body: A model study. Phys. Rev. B 62, 570-578 (2000)

[33] Kiselev, S. I. et al. Microwave oscillations of a nanomagnet driven by a spin-polarized current. *Nature* 425, 380-383 (2003).

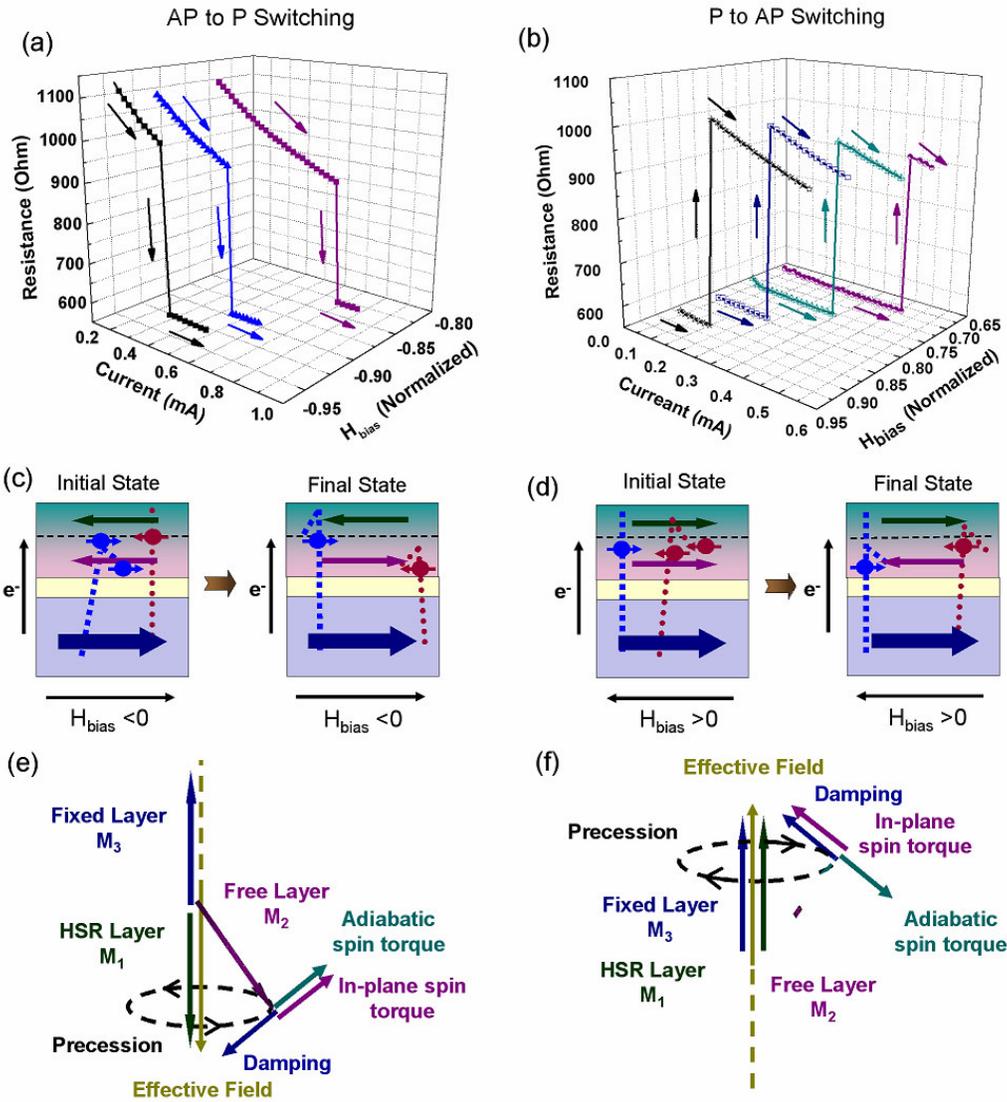

Fig.1: The transport behavior of the MTJs with a hybrid free layer structure for AP to P switch (a,c,e) and P to AP switch (b,d,f). Experimental results are shown in (a) and (b). The schematic drawings of the initial state and the final state of MTJ devices are shown in (c) and (d). The drawings of the motion of the free layer moment are shown in (e) and (f).

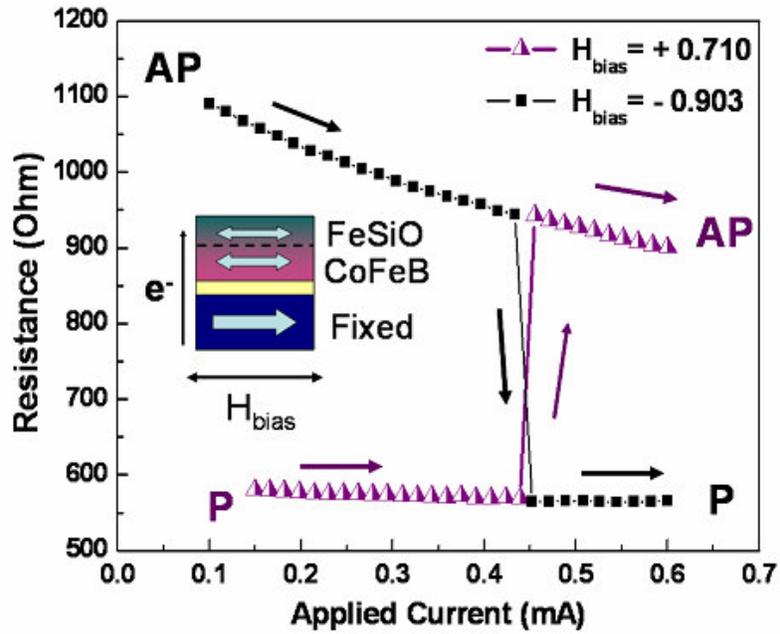

Fig.2: Single-polar current induced switching of the free layer of the MTJ devices with a hybrid free layer structure in both directions under different bias field. (Black square curve shows AP to P switch, and purple triangle curve shows P to AP switch.)

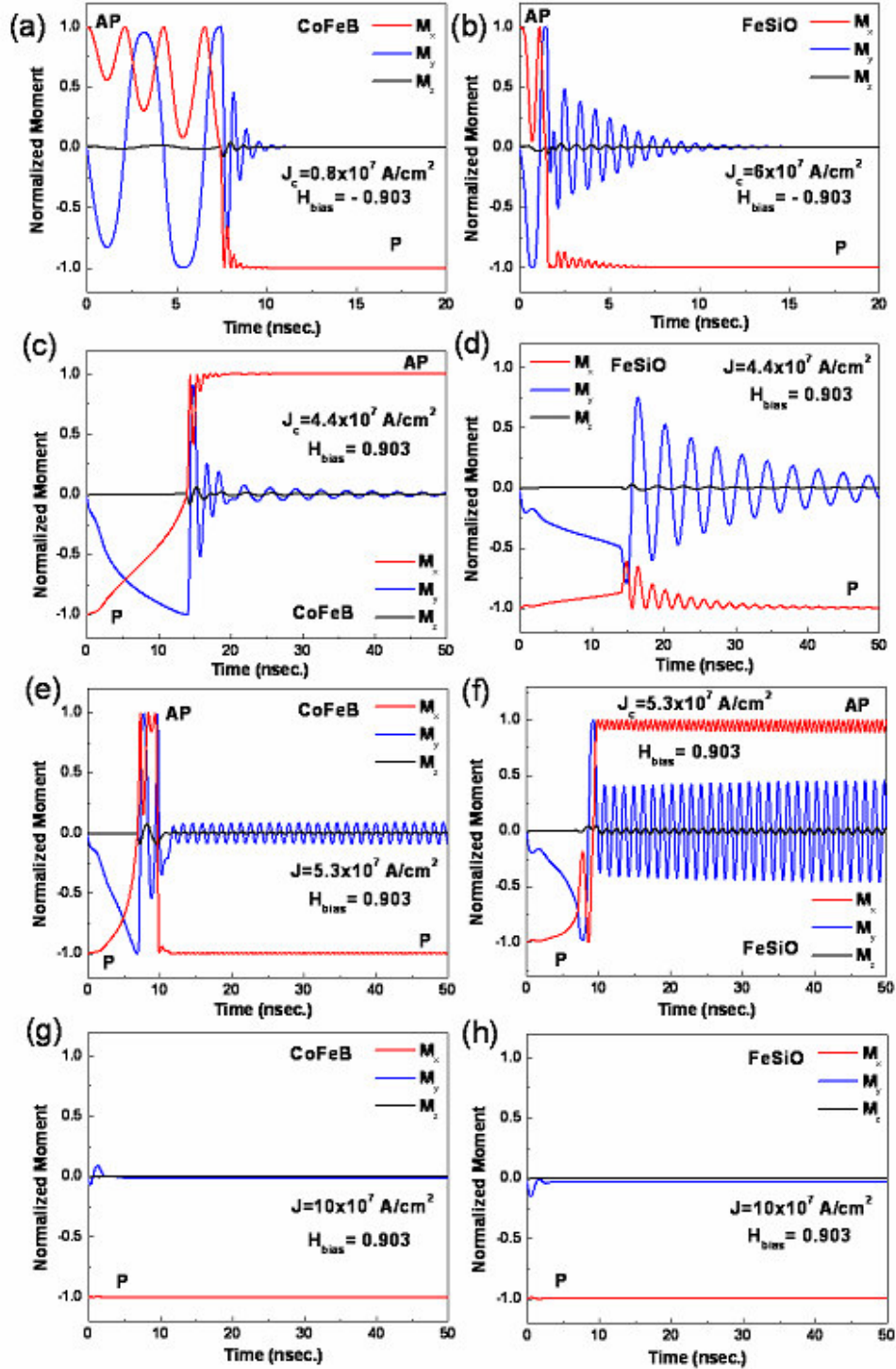

Fig.3: Simulation results of AP to P switching (a, b) and P to AP switching (c~h). The transport behavior of the CoFeB free layer is shown in (a, c, e, g) and that of the FeSiO high spin reflection layer is shown in (b,d,f,h). The applied current increases from (c) to (g) and also from (d) to (h). X,Y, and Z direction represent the in-plane easy-axis, in-plane hard-axis, and out-of-plane direction, respectively.

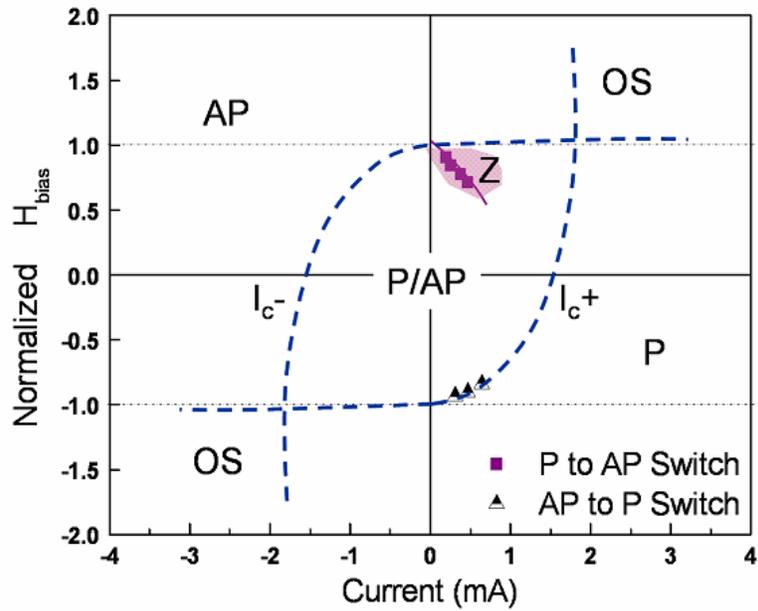

Fig.4: The possible I-H phase diagram for the switching of MTJ devices. Dashed line shows the phase boundaries of conventional MTJs, Square and Triangle dots are experimental data for the MTJ devices with hybrid free layer structures. Unique P to AP switching with a positive current forms a special region named zone Z.

Supplementary Information

The CoFeB and FeSiO thin films were analyzed in order to confirm the criteria to realize giant adiabatic spin torque as mentioned in the full paper. The analyzed film structures for CoFeB and FeSiO are: Si sub/$SiO_2$ (100 nm)/ CoFeB (3~5 nm)/ Ta (2 nm) and Si sub/$SiO_2$ (100 nm) /FeSiO (5 nm)/ Ta (2 nm), respectively. The samples were annealed at 310º C for 2 hrs with a 4 kOe field.

The composition and phase analysis of the FeSiO film was done by x-ray photoelectron spectroscopy (XPS). Fig1s(a) shows the Fe 2p3 high resolution scan spectrum. The peak position is 706.4 eV, which is close to the standard Fe 2p3 peak of 706.75 eV[1]. There is no peak at 710.7 eV, which belongs to $Fe_2O_3$ phase. Fig.1s(b) shows the Si 2p high resolution scan. The two peaks are 103.17 eV and 99.08 eV, which represent $SiO_2$ and FeSi or Si phases, respectively. The area ratio of these two peaks is 1:1. Since the XPS can detect up to 10 nm depth below the surface, some of the $SiO_2$ signal may come from $SiO_2$ layer below the FeSiO layer in the sample. The FeSi peak (~99 eV) and the Si peak (99.15 eV) are very close to each other. Thus, it is difficult to distinguish the FeSi and Si phases from XPS result. However, the transport results show that giant adiabatic spin torque is generated by inserting the FeSiO layer adjacent to the CoFeB free layer, which indicates that the FeSiO layer has a high spin polarization ratio. Therefore, it is believed that $Fe_3Si$ phase may be formed in the FeSiO annealed samples.

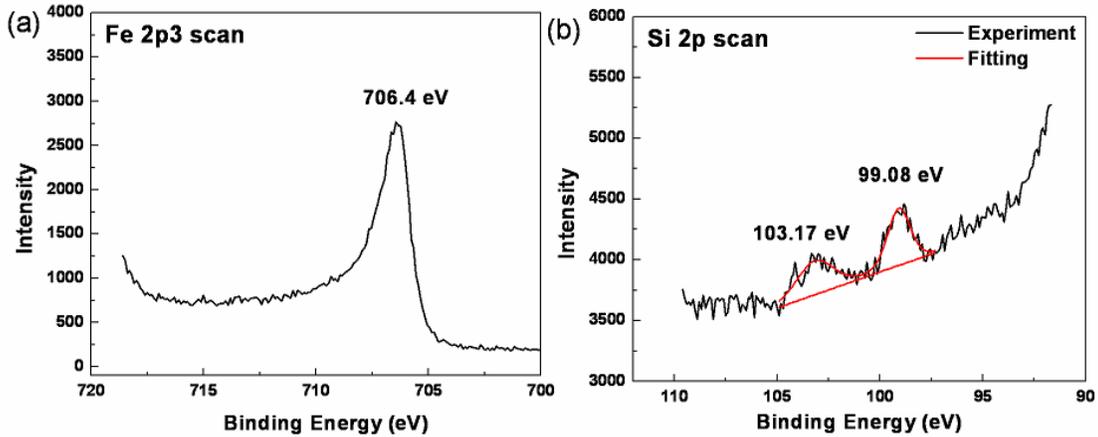

Fig.1s X-ray photoelectron spectroscopy (XPS) results of (a) Fe 2p3 scan and (b) Si 2p scan.

The damping constant was measured by the ferromagnetic resonance (FMR) method as shown in Fig2s. The damping constants of the FeSiO film and the CoFeB film are estimated as 0.015 and 0.008, respectively. Since the measured film is ultra thin (3~5 nm) with good uniformity, we assume that two–magnon scattering is small and the degree of inhomogeneity is low. Therefore, we assume that these measured damping constants are close to the intrinsic damping constants for CoFeB and FeSiO, which is used in the numerical simulations.

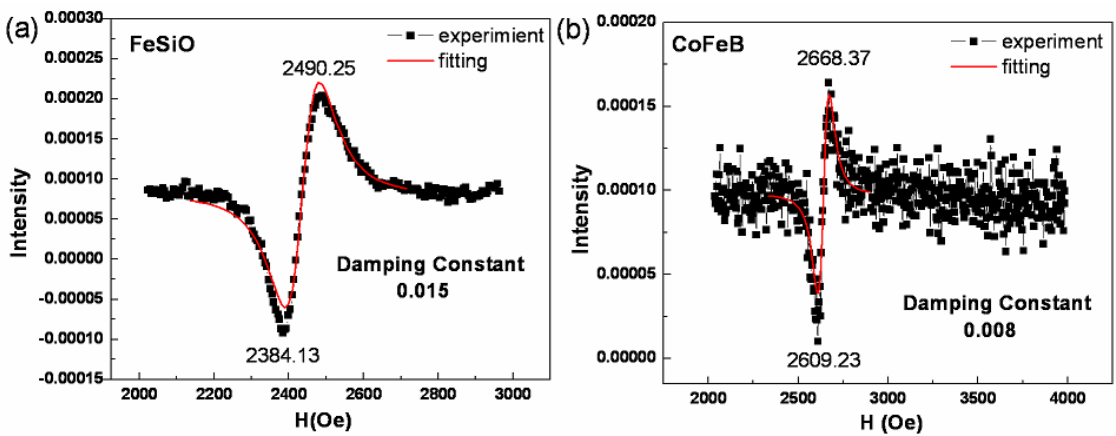

Fig.2s The ferromagnetic resonance (FMR) spectra (black curves) for 3nm CoFeB film (a) and 5nm FeSiO film (b). The red curves are the fitting curves.

Fig.3(a) shows the resistance-field loop for MTJ devices with hybrid free layer structures. The TMR ratio is 98% with an RA value at parallel state of 11.6 $\Omega \times \mu m^2$. In order to exclude the thermal heating effect for the unique parallel (P) to antiparallel (AP) switching, negative current is also applied as shown in Fig.3s(b). When the negative applied current is three times larger than the positive switching current, the devices still remain in the P state. Multiple scans have been done under the same measurement conditions, as shown in the multiple curves with the negative current. Therefore, it is believed that the giant adiabatic spin torque is indispensable for the unique P to AP switching with positive current, and is not due to the thermal heating effect.

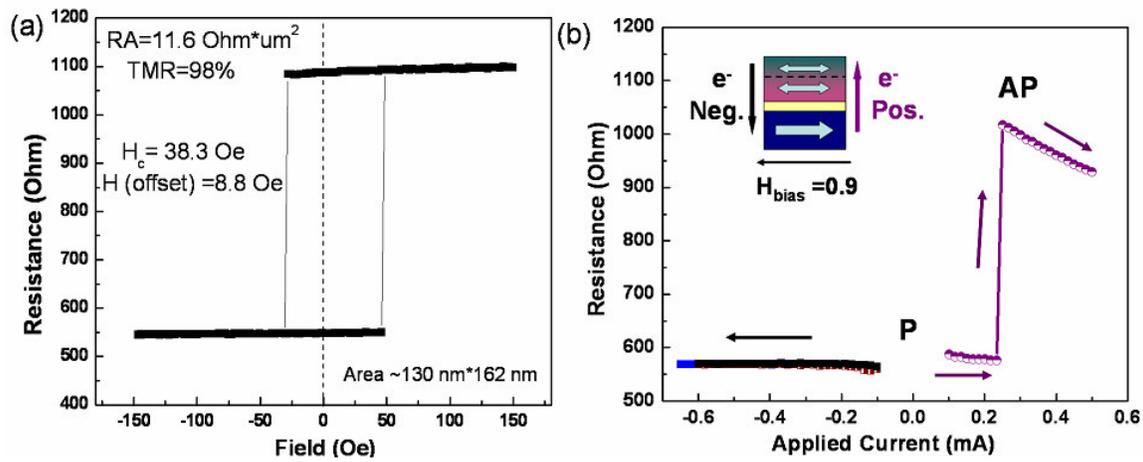

Fig.3s(a) The resistance-field loop of MTJ devices with hybrid free layer structures. (b) The measurement curves of MTJ devices with hybrid free layer with initial parallel state under both positive and negative current in order to exclude the thermal heating effect.

---

[1] Wagner, C.D., Riggs, W.M., Davis, L.E., Moulder, J.F., & Muilenberg, G.E. (Eds.), Handbook of X-ray photoelectron spectroscopy. (Perkin-Elmer, Eden Prairie, MN. 1979)